\documentclass[twocolumn]{article}

\usepackage[a4paper, margin=0.6in]{geometry}
\usepackage{tikz}
\usetikzlibrary{shapes.geometric, arrows.meta, positioning}
\usepackage{amsmath}
\usepackage{amsfonts}
\usepackage{siunitx}
\usepackage{graphicx}
\usepackage{xcolor}
\usepackage{dcolumn}
\usepackage{bm}
\usepackage[utf8]{inputenc}
\usepackage[T1]{fontenc}
\usepackage[sorting=none, url=false]{biblatex}
\usepackage{mathptmx}
\usepackage{etoolbox}
\usepackage{csquotes}
\usepackage{hyperref}
\usepackage{subcaption}
\usepackage{tcolorbox}
\usepackage{booktabs}
\usepackage{cleveref}

\graphicspath{ {figures/} }

\newcommand{\cov}[2][]{\text{cov}_{#1} \left[ #2 \right]}
\newcommand{\var}[2][]{\text{var}_{#1} \left[ #2 \right]}
\newcommand{\E}[2][]{\mathbb{E}_{#1} \left[ #2 \right]}

\title{Implementation and Extension of the Variance-Reduced BGK Method in PICLas}
\author{
L. Teichröb\footnote{Institute of Space Systems, University of Stuttgart.}\hspace{2mm}\footnote{Corresponding author. \url{Leon.Teichroeb@protonmail.com}} \and F. Garmirian\textsuperscript{*} \and M. Pfeiffer\textsuperscript{*}}%
\date{June 24, 2026}

\begin{document}

\maketitle

\begin{abstract}
Traditional particle-based kinetic methods, such as DSMC, suffer from prohibitive computational cost in low-signal flows, where the deviation from thermodynamic equilibrium is small and statistical noise overwhelms the signal of interest.
The Variance-Reduced BGK-DSMC scheme is further advanced and implemented to support this class of flows in the open-source gas-kinetics framework \mbox{PICLas}.
Modified versions of flow estimators and collision operators enhancing stability are developed.
The Shakhov and Ellipsoidal Statistical models for BGK are demonstrated, along with entirely new features such as adaptive equilibria, variable particle weights and domain axisymmetry.
The implementation is validated using synthetic benchmarks, 1D, 2D and axisymmetric simulations.
Comparison of VRBGK to BGK simulations shows exact agreement of the models.
A further comparison with an analytical solution of thermal transpiration in a microchannel showcases the low-signal efficiency of the method as well as newly proposed features.
\end{abstract}

\section{Introduction}
\label{sec:introduction}

\message{The text column width is \the\linewidth}

\begin{figure}
  \centering
  \includegraphics[width=0.6\linewidth]{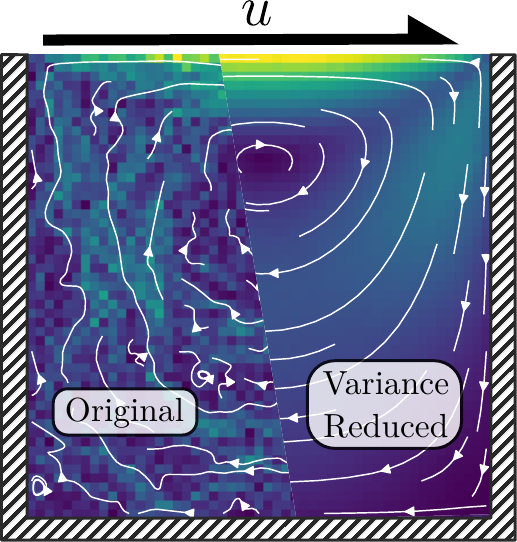}
  \caption{Comparison of the velocity field of a noisy (left) and a variance-reduced (right) Monte Carlo lid-driven cavity simulation.}
  \label{fig:vrbgk_vs_bgk}
\end{figure}

Particle simulation methods are a widespread solution for the investigation of rarefied gas flows, with direct simulation Monte Carlo (DSMC)~\cite{BirdDSMC} as the standard tool for high altitude aerodynamics~\cite{lebeau2001application,tumuklu2018unsteadiness,pfeiffer2024numerical} or microscale gas dynamics~\cite{oran1998direct,alexeenko2002numerical,akhlaghi2023comprehensive}.
However, the stochastic nature of the Monte Carlo method necessarily leads to non-physical statistical noise in the results, as illustrated in Figure \ref{fig:vrbgk_vs_bgk}.
The quantities of interest, such as flow velocity or temperature, are estimated by sampling the position and velocity of simulation particles, thus leading to a variance in the results that scales inversely with the number of samples.
For slow rarefied flows, such as those occurring in microelectromechanical systems (MEMS), the signal-to-noise ratio can become vanishingly small.
Maintaining a constant signal-to-noise ratio (SNR) requires the sample size to scale inversely with the signal magnitude $\sqrt{N} \propto 1/\left| x \right|$; for small signal flows this quickly leads to prohibitive computational costs.

Another class of methods solves rarefied gas problems deterministically by discretizing the Boltzmann equation, or its Bhatnagar-Gross-Krook (BGK) approximation~\cite{bhatnagar1954model}, in phase space.
Such discrete velocity methods (DVM)~\cite{mieussens2000discrete,xu2010unified,guo2013discrete,felix_dvm} are free of statistical noise, but can require a fine discretization of velocity space to accurately capture the distribution function.
In particular for three-dimensional cases, solving the differential equation on the fine six-dimensional space-velocity grid again leads to high computational costs.
Furthermore, the adaptability of particle methods to boundary conditions and velocity distributions is a clear advantage over DVM.

Therefore, several techniques have been developed to reduce the variance in DSMC simulations, for instance deviational formulations of the Boltzmann equation~\cite{og_deviational_lvdsmc} or the usage of control variates~\cite{VRPaper,AlMohssen2010AnEW}.
Because these methods retain the restrictive requirements of DSMC in terms of spatial and temporal discretization, such noise-reduced particle methods have also been developed for the BGK~\cite{VRBGKPaper,dsbgk} and the Fokker-Planck~\cite{gorji2015variance,netterdon2026variance} collision models.
Besides the variance reduction, these approaches solve approximations of the Boltzmann equation without having to resolve mean free paths or collision frequencies.

The Information Preservation (IP) method, first proposed by J. Fan and C. Shen~\cite{FAN2001393}, is suitable for the application to low-signal, pressure-driven Poiseuille flows, such as those found in Knudsen pumps~\cite{SHEN2003512}.
Later, the method was successfully extended to also correctly model the effects arising from temperature gradients~\cite{10.1063/1.1407562, SUN2002400}, including thermal transpiration~\cite{ZHANG20117250}.

Among these noise reduction techniques, the variance-reduced BGK (VRBGK) method has been selected for extension and application within the scope of this work, since its formulation allows for a relatively straightforward implementation into an existing BGK particle solver.
This method employs a control variate approach first introduced by \textcite{AlMohssen2010AnEW} as variance-reduced DSMC (VRDSMC) and adapted for the BGK model by \textcite{VRBGKPaper}.
Contrary to the deviational formulation used by \textcite{og_deviational_lvdsmc}, the importance weights associated with particles remain positive, which is a desirable property for the inclusion of complex processes in the simulated gas.
The VRDSMC class of methods provide good performance and an ideal foundation for inclusion of more advanced solver features, due to their greater simplicity both in theoretical correctness and implementation.
In addition, the conditional probability generalization of VRDSMC offers a reliable path forward for the inclusion of new models.

In this paper, we present a novel implementation of the BGK variation of VRDSMC into the open-source gas-kinetics framework \mbox{PICLas}~\cite{fasoulas2019combining}, with a focus on a rigorous definition of the importance weights and corresponding estimators to improve the overall stability and variance reduction.
The new implementation is feature-rich and aims to make the VRBGK method accessible to researchers for direct application to practical problems.
The implementation supports efficient 2D axisymmetric simulations, 3D, adaptive boundaries and numerous other quality-of-life features offered by building on top of a well-established solver.
Additionally, the method is extended to support the Shakhov (SBGK)~\cite{Shakov} and the ellipsoidal statistical (ESBGK)~\cite{esbgk} collision models, which, unlike the classic BGK model, attain the correct Prandtl number in simulations.
By extending the local equilibrium approach, simulations encompassing regions of differing densities and temperatures become accessible to VRBGK.


Section \ref{sec:vr_formulation} begins with the theory underlying control variates as well as a proof of the small-signal variance scaling to provide a foundation for discussing the changes we propose to the original method.
This includes both a modification to the calculation of flow variables and to the relaxation operator, enhancing their stability.
Practical questions of implementation, such as particle cloning, boundary-handling, and adaptive equilibria are documented in Section \ref{sec:implementation}.
Validation of the method using a combination of synthetic benchmarks as well as practical simulations is presented in Section 4 and is followed by concluding remarks on difficulties and opportunities facing the class of VRDSMC methods.

\section{Method}
\label{sec:vr_formulation}

Variance reduction in \mbox{PICLas}~\cite{PiclasPaper} is achieved using the BGK variation of the VRDSMC method, (VR)\textsuperscript{2}-BGK~\cite{VRBGKPaper}, with a number of important modifications that are detailed in the following section.
\enquote{(VR)\textsuperscript{2}-BGK} will be referred to as VRBGK for the remainder of this text.

\subsection{Control Variate Theory}
For DSMC methods, a variance reduction method seeks to reduce the variance of the estimator of a velocity moment $R(c) = c^k$, $k \in \mathbb{N}_0$ without increasing the required number of samples.
The \enquote{control variate} method is well known and has been in use since the inception of Monte Carlo methods for neutron transport problems~\cite{Lavenberg_ControlVariates}.
The control variate method works on the principle of finding a correlated estimate $\bar{R}_W$ with analytically known mean $\E{R_W}$, allowing the common statistical noise to be removed. 
Taking $\beta$ to be any real constant,
\begin{align}
    \bar{R}_{VR} &= \bar{R} - \beta \left( \bar{R}_W - \E{R_W} \right) \label{eq:control_variates}
\end{align}
is an unbiased estimator of $R$ that, for a suitable choice of $\beta$ and correlation of the control variate, benefits from reduced variance.
In the context of this work, due to the near-unity correlation $corr(R, R_W) \approx 1$ that is observed, $\beta=1$ is selected and is therefore omitted hereafter.

The VRBGK and VRDSMC methods propose a prediction of thermodynamic equilibrium as a choice of control variate, co-simulated on the set of original particles to maximize correlation.
This is a plausible choice, since the moments of thermodynamic equilibrium are known analytically.
By introducing a per-particle weight and suitable update rules, it becomes possible to simultaneously simulate both the control and the problem at minimal additional cost. 
This procedure is, in form, equivalent to importance sampling but aims to minimize the variance of $\bar{R}_{VR}$ as opposed to $R_W$ directly.
For this reason, we will refer to these weights as Variance Reduction (VR) weights, instead of importance weights.

Statistically, expressing the equilibrium as weighted samples of a different distribution is a form of probability space transformation where the weight is the Radon-Nikodym derivative.
The distributions of interest are the solution $f$ and an arbitrary, Maxwell-Boltzmann distributed equilibrium state $f_{eq}$.
Because it can be determined locally without considering the entire simulation domain, it is favorable to work with the particle number density $F(\bm x, \bm c) = N f(\bm x, \bm c)$ rather than $f$ directly.
Here, $N$ denotes the number of physical particles contained in the domain of $f$.
Beginning with the expectation of an arbitrary velocity moment $R$, the appropriate VR weight arises naturally from the transformation from $F_{eq}$ to the actual distribution $F$: 
\begin{align}
    \E[f_{eq}]{R} &= \frac{\iint  R(\bm c) F_{eq}(\bm x, \bm c) d\bm x \, d\bm c}{\iint F_{eq}(\bm x, \bm c) d\bm x \, d\bm c} \notag \\
                  &= \frac{1}{N_{eq}} \iint  R(\bm c) \frac{F_{eq}(\bm x, \bm c)}{F(\bm x, \bm c)} F(\bm x, \bm c) d\bm x \, d\bm c \notag \\
    \intertext{Next, we define the particle \enquote{VR-weight} $W$ as $F_{eq} / F$ and normalize by $N = \iint F d\bm x \, d\bm c$, yielding the transformed expectation:}
    \E[f_{eq}]{R} &= \frac{N}{N_{eq}} \frac{\iint R(\bm c) W(\bm x, \bm c) F(\bm x, \bm c) d\bm x \, d\bm c}{\iint F(\bm x, \bm c) d\bm x \, d\bm c} \notag \\
                  &= \frac{N}{N_{eq}} \E[f]{RW} \label{eq:equilibrium_expectation}
\end{align}
This expectation is expressed over the probability space $f$ as required.
The interpretation is that the expectation value of weighted samples from the distribution $f$ is that of the equilibrium moment, scaled by the ratio of the densities.
This result applies to any subset of the domain volume and is therefore also true for individual simulation cells.
Inserting Eq. \ref{eq:equilibrium_expectation} into the control variate scheme of Eq. \ref{eq:control_variates},
\begin{equation}
  \overline{R}_{VR} = \overline{R} - \overline{RW} + \frac{N_{eq}}{N} \E[f_{eq}]{R}, \quad W = \frac{F_{eq}}{F}
  \label{eq:vr_expectation}
\end{equation}
we obtain a variance reduced estimator for any velocity moment $R$.

By its definition, $W$ is a scalar field over the particle state space and depends on knowledge of the solution $F$.
Therefore, for practical problems, calculating it exactly is not possible.
However, the rules dictating the per-particle, Lagrangian evolution of $W$ are quite simple, both computationally and conceptually.
Because the only values of $W$ required to estimate $\E[f]{RW}$ will be those corresponding to actual particles, it will be sufficient to track $W$ as a per-particle weight.

Finally, to complete the theoretical foundation, we will show that the coefficient of variation of Eq. \ref{eq:vr_expectation} is finite and constant for arbitrarily small signals.
This was shown for the estimation of velocity in the presence of velocity deviations by \textcite{AlMohssen2010AnEW}.
Here it will be shown for arbitrary moments and include combined deviations.
Despite $F$ not being Maxwell-Boltzmann distributed, processes such as initialization, relaxation, and diffuse wall interactions are.
For classic VRBGK, this analysis is valid for every particle VR-weight in the domain, as weights are only set in these three scenarios. 
We use the temperature $T$ and bulk velocity $u$ of any one of these distributions to set an upper bound on the deviation from equilibrium.
If, by suitable choice of reference frame, the equilibrium bulk velocity is $u_{eq} = 0$, all possible deviations of the Maxwell-Boltzmann distribution are $\epsilon_u = u$, $\epsilon_T = \frac{T - T_{eq}}{T_{eq}}$, and $\epsilon_N = \frac{N - N_{eq}}{N_{eq}}$.
Expressing $F$ in terms of deviations from $F_{eq}$:
\begin{equation*}
  F = \frac{N_{eq} (1 + \epsilon_N)}{\sqrt{2 \pi \sigma_{eq}^2 (1 + \epsilon_T)}} \exp \left[ -\frac{(c - \epsilon_u)^2}{2 \sigma_{eq}^2 (1 + \epsilon_T)} \right]
\end{equation*} 
allows us to derive the linear perturbation of $F$ around $F_{eq}$.
The individual deviations are summarized by their geometric mean $\epsilon = \sqrt{\epsilon_u^2 + \epsilon_T^2 + \epsilon_N^2}$.
We assume small deviations $\epsilon \leq 1$, and linearize first around the PDF $f$:
\begin{align}
  f &= f_{eq} + \frac{\partial f}{\partial \epsilon_u} \epsilon_u + \frac{\partial f}{\partial \epsilon_T} \epsilon_T + \mathcal{O}(\epsilon^2) \notag \\
    &= f_{eq}\left( 1 + \frac{c\epsilon_u}{\sigma_{eq}^2} + \frac{c^2\epsilon_T}{2 \sigma_{eq}^2} - \frac{\epsilon_T}{2} \right) + \mathcal{O}(\epsilon^2) \notag
\end{align}
Next, the density term $N_{eq} (1 + \epsilon_N)$ is reintroduced:
\begin{equation*}
  F = F_{eq}\left( 1 + \frac{c\epsilon_u}{\sigma_{eq}^2} + \frac{c^2\epsilon_T}{2 \sigma_{eq}^2} - \frac{\epsilon_T}{2} + \epsilon_N \right) + \mathcal{O}(\epsilon^2)
\end{equation*} 
This can be inserted into the definition of the density-free particle weight (Eq. \ref{eq:vr_expectation}) to obtain the approximation
\begin{equation*}
  1 - W = \mathcal{O}(\epsilon),
\end{equation*}
allowing the variance of an arbitrary velocity moment $\var{(1 - W)c^n}$ to be bounded:
\begin{align}
  \var[f]{(1-W)c^n} &\leq \E[f]{((1-W)c^n)^2} \notag\\
                    &= \mathcal{O}(\epsilon^2) \E[f]{c^{2n}} \notag
\end{align}
The term $\E[f]{c^{2n}}$ is approximately equal to the constant term $(1 + \mathcal{O}(\epsilon))\E[f_{eq}]{c^{2n}}$ and does not influence the order.
Therefore, the standard deviation of any variance-reduced velocity moment scales linearly with the total deviation from equilibrium:
\begin{equation*}
  \sigma_f \left[ (1-W)R \right] = \mathcal{O}(\epsilon)
\end{equation*}
In the common scenario of isolated deviations in velocity or temperature, their respective coefficients of variation are:
\begin{gather}
  \frac{\sigma_u}{u} = \frac{\sqrt{\mathcal{O}(\epsilon^2)}}{\mathcal{O}(\epsilon)} = \mathcal{O}(1) \label{eq:u_variance_magnitude}\\
  \frac{\sigma_T}{T - T_{eq}} = \frac{\sqrt{\mathcal{O}(\epsilon^2)}}{\mathcal{O}(\epsilon)} = \mathcal{O}(1) \label{eq:T_variance_magnitude}
\end{gather}
Therefore, VRBGK offers efficient scaling of the simulation to any signal magnitude and for all velocity moments.
It should be restated that the level of variance reduction across all moments is dependent on the \textit{total} deviation $\epsilon$, meaning that non-vanishing deviation in one quantity leads to non-vanishing variance in every other.

\subsection{Realization of Estimators}
In the density correction term of Eq. \ref{eq:vr_expectation}, $N_{eq}$ is known, since the equilibrium state is prescribed, but estimating the unknown quantity $N$ poses a challenge.
Classical control variate theory dictates that the expectation value (i.e. the final term in Eq \ref{eq:vr_expectation}) is to be known analytically~\cite{ROSS2013153}, otherwise it may introduce bias or variance.
In practice, obtaining the analytical value is impossible, since perfect knowledge of the sought-after solution would be required.
Instead, a numerical approximation $\bar{N} \approx N$ with low variance and minimal bias is required.

There is some ambiguity in the resolution of this term.
Indeed, there are differences between the works on VRDSMC and VRBGK, as well as in the current \mbox{PICLas} implementation.

In the VRDSMC scheme presented by \textcite{VRPaper}, a simple counting of the particles is used, $\bar{N} = \mu N_{sim}$, where $N_{sim}$ is the current number of simulation particles in the cell and $\mu$ is the \enquote{particle factor} describing the number of real particles represented by each simulation particle.
For non-zero $\E[f_{eq}]{R}$, this formulation introduces Poisson-noise in the final term of Eq. \ref{eq:vr_expectation}.
Because $\E[f_{eq}]{c^n} = 0$ for $n \neq 2$, this term only affects estimation of the second velocity moment.
This noise does not scale with proximity to equilibrium, and thus disrupts the desired small signal scaling $\sigma \propto \mathcal{O}(\E{R} - \E[f_{eq}]{R})$.
This is confirmed numerically in Section \ref{sec:synthetic_tests}.

The VRBGK scheme by \textcite{VRBGKPaper}, arising from the work on VRDSMC, instead uses the variance reduced estimator $\bar{N}_{VR}$ of the particle number
\begin{equation}
  \bar{N}_{VR} = \mu \sum^{N_{sim}}_{j=1} (1 - W_j) + N_{eq}
  \label{eq:N_vr}
\end{equation}
which benefits from diminishing variance as equilibrium is approached.
As such, this approximation produces the correct small-signal scaling.
However, this approach suffers from two sources of bias that can serve to destabilize the simulation.

The first fundamental source of bias arises directly from the mathematical properties of applying the inverse to a random variable.
By taking the expectation value over the random variable $\bar{N}_{VR}$, and recognizing that $f(x) = 1/x$ is a strictly convex function for $x > 0$, a lower bound can be established using Jensen's inequality:
\begin{equation*}
  \E{\frac{1}{\bar{N}_{VR}}} \ge \frac{1}{\E{ \bar{N}_{VR} }}
\end{equation*}
Using the fact that the sum of VR-weights corresponds to the equilibrium particle number, $\mu \E{\sum W} = N_{eq}$, the expectation of the variance-reduced number simplifies to $\E{\bar{N}_{VR}} = N$.
This yields:
\begin{equation*}
  \E{\frac{1}{\bar{N}_{VR}}} \ge \frac{1}{N}
\end{equation*}
Because $\bar{N}_{VR}$ has variance and is not constant for any simulation of practical interest, the strict convexity of the inverse function strengthens Jensen's inequality to show that this scheme overestimates the fraction:
\begin{equation*}
  \E{\frac{1}{\bar{N}_{VR}}} > \frac{1}{N}
\end{equation*}

The second source of bias arises from reliance on the identity $\mu \E{\sum W} = N_{eq}$.
This expectation holds at open boundaries and during initialization of the simulation, however when particles are relaxed on the basis of $\bar{N}_{VR}$, the VR-weight $W$ is no longer a martingale.
If we consider the sum-form of $\bar{N}_{VR}$:
\begin{equation*}
  \bar{N}_{VR} = N - \mu \sum W + N_{eq}
\end{equation*}
it is evident that $\cov{\bar{N}_{VR}, \sum W} < 0$.
When the relaxation operator scales the drawn weights by a factor $W^\prime \propto 1/\bar{N}_{VR}$ the correlation $\cov{W^\prime, \sum W} > 0$ causes small statistical fluctuations to become unstable.

For these reasons, in place of the existing expressions that use $N$ and $\bar{N}_{VR}$, we propose a new, simple unbiased approximation term.
Taking the entire quotient from Eq. \ref{eq:vr_expectation}, we show that it is equal to the expectation value of the VR-weights:
\begin{align}
  \frac{N_{eq}}{N} &= \frac{\iint F_{eq} d\bm x \, d\bm c}{\iint F d\bm x \, d\bm c}
             = \frac{\iint W F d\bm x \, d\bm c}{\iint F d\bm x \, d\bm c} \notag \\
             &= \E{W} \label{eq:unbiased_expectation_part} \notag
\end{align}
Estimating $\E{W}$ is preferable, since it may be done using the unbiased weighted sample mean,
\begin{equation}
  \bar{W} = \frac{\sum \mu_j W_j}{{\sum \mu_j}}
  \label{eq:unbiased_expectation_estimator}
\end{equation}
Moreover, it continues to be unbiased even in the presence of deviations in the weight sum.
This is particularly advantageous for the relaxation operator, which only becomes marginally stable if there is no positive feedback present in the bias.
Section \ref{sec:synthetic_tests} provides numerical results supporting the choice of Eq. \ref{eq:unbiased_expectation_estimator} over prior methods.

To obtain the variance-reduced estimator as it is used in \mbox{PICLas}, we permit the particle factor to vary per-particle and the sample mean is substituted for $\overline{R}$ and $\overline{RW}$ in Eq. \ref{eq:vr_expectation}:
\begin{equation}
  \bar{R}_{VR} = \frac{\sum \mu_j (1 - W_j) R(c_j)}{\sum \mu_j} + \frac{\sum \mu_j W_j}{{\sum \mu_j}} \E[f_{eq}]{R}
  \label{eq:R_vr}
\end{equation}
where $c_j$, $W_j$, and $\mu_j$ are the velocity, VR-weight and particle factor of the sample $j$.
This expression is equally valid on any subset of the domain, i.e. any computational cell.

Assuming constant $\mu_j$ for comparison, the first term corresponds precisely to the VRDSMC estimators used by Al-Mohssen~\cite{AlMohssen2010AnEW}, but with $\bar{W}$ used in place of $N_{eq} / \bar{N}$:
\begin{equation}
  \bar{R}_{VR} = \frac{1}{N} \sum \mu (1 - W_j) R(c_j) + \frac{N_{eq}}{N} \E[f_{eq}]{R}
  \label{eq:almohssen_estimator}
\end{equation}
As discussed earlier, the VRDSMC estimators do not offer small-signal scaling for non-zero equilibrium moments for this reason. 
Landon~\cite{VRBGKPaper, VRThesis} presents the VRBGK method utilizing estimators using constant particle factors and the variance-reduced sample size $\bar{N}_{VR}$ (Eq. \ref{eq:N_vr}):
\begin{equation}
  \bar{R}_{VR} = \frac{1}{\bar{N}_{VR}} \sum \mu (1 - W_j) R(c_j) + \frac{N_{eq}}{\bar{N}_{VR}} \E[f_{eq}]{R}
  \label{eq:landon_estimator}
\end{equation}
As discussed, this solves the variance issue affecting the VRDSMC scheme, but leads to instability and bias in certain cases.

The use of $\bar{N}_{VR}$ to normalize the sum is also of interest. 
Ostensibly, the choice to utilize a variance-reduced sample size aims to reduce contributions of the random, Poisson-distributed cell particle number~\cite[p.9]{BirdDSMC} to the estimator variance. 
However, due to the scaling of the sum with the varying sample size $N$, normalizing by $\bar{N}_{VR}$ serves to increase the variance rather than to diminish it (Section \ref{sec:synthetic_tests}).
Because the magnitude of the sum also scales with $(1 - W_j)$, VRBGK, as it is presented by Landon, still yields scaling of the order $\sigma \propto \mathcal{O}(\E{R} - \E[f_{eq}]{R})$, thus providing the algorithmic advantage for small-signals.

\subsection{Relaxation Models}

The weight rule for BGK relaxation stems from Landon~\cite{VRPaper}.
It can be derived by substituting $F_{eq} = WF$ and assuming that the time derivative of the BGK collision rule vanishes at equilibrium.
Taking $F_{T}$ as the BGK target distribution and $\nu$ as the relaxation rate,
\begin{align*}
    0 = \left( \frac{\partial F_{eq}}{\partial t} \right)_{\text{coll}} &= W \left( \frac{\partial F}{\partial t} \right)_{\text{coll}} + \left( \frac{\partial W}{\partial t} \right)_{\text{coll}} F \\
    &= W \nu \left( F_{T} - F \right) + \left( \frac{\partial W}{\partial t} \right)_{\text{coll}} F
\end{align*}
Solving for the change of weights in time and introducing the relaxed weight $W_T = F_{eq} / F_T$:
\begin{equation}
  \left( \frac{\partial W}{\partial t} \right)_{\text{coll}} = \frac{F_{T}}{F} \nu \left( W_T - W \right)
  \label{eq:weight_relaxation_derivative}
\end{equation}
In~\cite{VRPaper}, validity is made conditional on $F_{T}$ being close to $F$, in which case $F_{T} / F \approx 1$, making the expression completely analogous to the BGK collision operator.
Therefore, mirroring the BGK approach, it is sufficient to set the weight of relaxing particles to $F_{eq}/F_{T}$ and to leave the weights of the rest of the particles in the simulation unchanged.
We propose that, despite differing in form from velocity relaxation, this scheme is exact even without assuming $F_{T} / F \approx 1$.
To illustrate, assume that for a given velocity $c$, $F_T$ is larger than $F$.
In this case, fewer particles of this velocity are removed than new ones are added, accentuating the otherwise linear relaxation towards $W_T$---the opposite occurs as $F_T$ becomes small.
This coincides precisely with the meaning of the term $F_T / F$ in Eq. \ref{eq:weight_relaxation_derivative}.

Alternatively, by taking the post-relaxation state to be a mix of relaxed and unrelaxed particles with $\alpha = 1 - e^{-\nu \Delta t}$:
\begin{align*}
  W' &= \frac{(1 - \alpha) F W + \alpha F_T W_T}{(1 - \alpha) F + \alpha F_T} \\
     &= \frac{(1 - \alpha) F_{eq} + \alpha F_{eq}}{F'} = \frac{F_{eq}}{F'}
\end{align*}
With $F'$ denoting the post-relaxation particle number density, the post-relaxation weights $W'$ are correct according to definition.

To assign post-relaxation weights to particles, the quotient $F_{eq}/F_T$ must be evaluated.
As discussed in the previous section, substituting $\bar{N}_{VR}$ for $N$ in the density ratio introduces bias.
Instead, \mbox{PICLas} employs the unbiased sample mean of the weights to scale the PDFs, yielding the relaxation scheme
\begin{equation*}
  W' = \frac{\sum \mu_j W_j}{\sum \mu_j} \frac{f_{eq}}{f_T(c')},
\end{equation*}
where the sum runs over the $N_{sim}$ particles in the cell.
Calculating the expectation of the post-relaxation weight $W^\prime$ over the post-relaxation velocity $c'$ first and then the random pre-relaxation particle ensemble $S$, we obtain:
\begin{align*}
  \E{W'} &= \E[(S, c')]{\frac{\sum \mu_j W_j}{\sum \mu_j} \frac{f_{eq}}{f_T(c')}} \\
    &= \E[S]{\frac{\sum \mu_j W_j}{\sum \mu_j}} = \E{W}
\end{align*}
As required, the mean weight is preserved.
While statistical fluctuations still cause random walk drift in the total weight, these are small enough to be readily corrected by weight conservation schemes.

The SBGK~\cite{Shakov} and ESBGK~\cite{esbgk} relaxation models were also implemented besides the standard BGK model~\cite{bhatnagar1954model}.
This only requires an adjustment of the weight assignment during relaxation to use the SBGK or ESBGK target distribution for $F_T$.
As mentioned earlier, \mbox{PICLas} implements the (VR)\textsuperscript{2}-BGK approach, meaning that $f_T$ is calculated using variance-reduced velocity and temperature.
When using SBGK or ESBGK, \mbox{PICLas} additionally computes the variance-reduced pressure and heat-flux tensor respectively.

\section{Implementation}
\label{sec:implementation}
One of the most advantageous features of VRBGK for variance-reduction in pre-existing codes is its seamless integration with BGK implementations.
The large majority of modifications merely add weight-handling code without modifying existing processes.
Specialized, variance-reduced versions of routines are rarely necessary, with the only notable exception being VR sampling routines.
The performance impact of these changes is also minimal, as revealed by the benchmarks in Section \ref{sec:thermal_creep}.
Since advection leaves particle VR-weights unchanged and typically only few particles undergo relaxation and/or boundary interactions, these do not significantly contribute to any changes in runtime.
Profiling indicates that the main performance driver in PICLas is the slightly more expensive moment calculation, as it involves every particle.
The final per-particle runtime cost is completely offset by the algorithmic small-signal advantage.

\paragraph{Adaptive Equilibrium}
When enabled, \mbox{PICLas} defines local VR-weights $W_{loc}$ according to individual, per-cell reference distributions $F_{eq,loc}$.
This concept is already employed by \textcite{VRPaper} to improve the efficiency of kernel density estimation required for VRDSMC.
While VRBGK does not utilize a kernel density estimation step, the use of variable equilibrium states can also be extended to further improve variance reduction.
The derivation of the VR-estimator made no assumptions regarding the choice of reference distribution and is entirely cell-local, therefore sampling is immediately compatible with the use of local-referenced weights.
Similarly, any arbitrary, local Maxwellian distribution remains unchanged under relaxation and is therefore a possible choice.
Since advection is derived under the assumption of a global equilibrium distribution, it instead requires weights to use a single, global reference.

For the implementation in \mbox{PICLas}, each particle retains only a single VR-weight that is transformed between local- and global-referenced modes.
These transformations are defined as:
\begin{equation}
  W = W_{loc} \frac{F_{eq}}{F_{eq,loc}} \, , \qquad W_{loc} = W \frac{F_{eq,loc}}{F_{eq}}
\end{equation}

As an estimate for the local equilibrium, \textcite{VRPaper} use the current timestep's variance-reduced state directly; however, it has been observed that for low particle numbers or moderate deviations from equilibrium, this no longer yields a stable scheme.
To alleviate this issue, \mbox{PICLas} uses the exponential moving average of estimations as local equilibria.
The hysteresis introduced by the moving average can lead to temporary increases in variance.
Should these become problematic for rapidly evolving flows, the smoothing factor---and consequently the convergence rate---can still be reduced through increased sample sizes as a secondary control mechanism.

Cell-local reference states are especially advantageous for simulations which are locally near-Maxwellian but that vary globally in density, velocity or temperature.
By using spatially varying equilibria, certain challenging simulations become possible or more efficient.
One such example is the long channel of a Knudsen pump, as simulated in a simplified form in Section \ref{sec:thermal_creep}.

\paragraph{Boundaries}
Reflecting, accommodating, and open boundary types can be used in combination with the presented VRBGK implementation.
Reflecting and accommodating boundaries for VR\-DSMC and VRBGK have already been demonstrated~\cite{AlMohssen2010AnEW,VRBGKPaper}; however, we give additional attention to the nuances of modelling double collisions.
Open boundaries have previously only been proposed~\cite{AlMohssen2010AnEW}. 
In addition, these have been extended to include adaptive subsonic boundaries according to \textcite{FARBAR201499}.
Intermediate accommodation coefficients are supported by varying the probability of reflecting versus accommodating wall-interactions.

On the basis of conditional probability arguments, it is clear that, for a given particle as our prior, the velocity after reflection off the wall boundary is deterministic and identical to that of the hypothetical equilibrium particle.
Therefore, the relative probability, i.e. VR-weight, does not change.
In contrast, for fully accommodating boundaries, the prior is largely irrelevant, as the particles are drawn from a wall-distribution independently of their incident velocity or other properties.
Therefore, explicit weight assignment is ideal and additionally serves to stabilize the weights.
Some information from the prior particles is still required to ensure that the density of the \enquote{emitted} particles matches the incident density.
One way of achieving this is to scale each particle's pre-collision VR-weight by the probability of the diffuse wall interaction:
\begin{equation}
  \label{eq:self_scaled_diffuse}
  W_i' = W_i \sqrt{\frac{T_{wall}}{T_{eq}}} \frac{f_{eq}(c_i)}{f_{wall}(c_i)}
\end{equation}
This is correct but contributes to decorrelation of the simulation over time.
Another method of density preservation~\cite{VRPaper} is to scale emitted particles by the mean weight of incoming particles within a timestep:
\begin{equation}
  \label{eq:mean_scaled_diffuse}
  W_i' = \frac{\sum \mu_j W_j}{\sum \mu_j} \sqrt{\frac{T_{wall}}{T_{eq}}} \frac{f_{eq}(c_i)}{f_{wall}(c_i)}
\end{equation}
While this method affords superior weight stability, it poses a challenge for particles that undergo multiple diffuse boundary interactions in a single timestep.
Particles must have a valid VR-weight for every boundary interaction.
However, for intermediate collisions, the post-collision weight according to Eq. \ref{eq:mean_scaled_diffuse} is not available because it depends on the mean over all particles.
This is especially prevalent in the corners of domains, where the fraction of particles undergoing double collisions in a short time span becomes significant.
These weights are essential and strongly influence the solution if omitted, especially in the presence of distinct wall temperatures.
To solve this issue, \mbox{PICLas} assigns particles a \textit{preliminary} VR-weight according to Eq. \ref{eq:self_scaled_diffuse}, making it valid for any consecutive interactions.
At the end of the advection step, every particle that experienced a diffuse boundary interaction is reassigned a new, stabilized weight by Eq. \ref{eq:mean_scaled_diffuse}.

The diffuse wall scheme is easily extended to support the use of local equilibrium states.
Assuming Eq. \ref{eq:self_scaled_diffuse} and Eq. \ref{eq:mean_scaled_diffuse} to be expressed in local quantities, a local-to-global transformation yields:
\begin{equation*}
  W_i' = W_{i,loc}' \frac{F_{eq}}{F_{eq,loc}} = \begin{cases}
    W_{i,loc} \sqrt{\frac{T_{wall}}{T_{eq,loc}}} \frac{f_{eq}(c_i)}{f_{wall}(c_i)} \frac{n}{n_{loc}} \\
    \frac{\sum \mu_j W_{j,loc}}{\sum \mu_j} \sqrt{\frac{T_{wall}}{T_{eq,loc}}} \frac{f_{eq}(c_i)}{f_{wall}(c_i)} \frac{n}{n_{loc}}
  \end{cases}
\end{equation*}
Typically, $f_{eq,loc}$ will be significantly closer to $f_{wall}$ than $f_{eq}$ and yield VR-weights and a solution with lower variance.
 
\mbox{PICLas} supports multiple types of open boundaries in combination with VRBGK.
As particle creation and deletion proceed according to a known distribution and independently of particle state, explicit weight assignment is a suitable method.
As such, all open boundaries stabilize the importance weights. 
Adaptive boundaries according to Farbar and Boyd~\cite{FARBAR201499} are implemented using variance-reduced sampling of the interior.
Critically, this entirely circumvents the hysteresis and time-inaccuracy that typically accompanies Farbar and Boyd's method, since no averaging of samples is required.

\paragraph{Variable Particle Factors}
\mbox{PICLas} includes support for variable particle factors---i.e. the number of real particles per simulation particle---in combination with VRBGK, which has not yet been demonstrated for this control variate scheme~\cite{VRPaper, VRBGKPaper}.
Variable particle factors are primarily used to more efficiently simulate large density fluctuations, multi-species, and axisymmetric flow.

To maintain a target factor $\mu(\bm x)$, \mbox{PICLas} stochastically duplicates or deletes particles based on their factor deviation from the target, resetting $\mu_i$ to the target value afterward. 
The essential observation is that this scheme assigns the same transition probabilities to a given particle, irrespective of any other simulation state.
Therefore, a given particle would have equal chances of being cloned or deleted in both a simulation of equilibrium or non-equilibrium,
\begin{align*}
  P_{eq}(\text{\enquote{Clone $i$}}|\mu_i, x_i) &= \\
  P(\text{\enquote{Clone $i$}}|\mu_i, \bm x_i) &= \frac{\mu_i}{\mu(\bm x_i)} - 1
\end{align*}
since the factor $\mu_i$ and position $\bm x_i$ of the particle $i$ are fixed priors, and the target factor $\mu$ is a function only of the particle position.
Therefore, the VR-weight of cloned or deleted particles remains unchanged according to the rule $W^\prime = W \frac{P_{eq}}{P}$.
This is an example of the use of the conditional probability interpretation to easily extend VRDSMC-type methods to new features.
It is of note that, like this example, many phenomena are purely determined by the individual particle's state and hence do not affect the VR-weight at all.
This result was also extrapolated to reveal that all types of symmetry transformation do not modify the particle VR-weights.

In the Section \ref{sec:thermal_creep}, axisymmetric flows that leverage variable particle factors are presented.

\section{Results}
\label{sec:results}

\begin{table}
  \centering
  \caption{
    Test matrix indicating which features of VRBGK are validated by each simulation.
    (Syn. = synthetic, Cou. = Couette, LDC = Lid-driven cavity, TT = Thermal Transpiration)}
  \label{tab:test_matrix}
  \begin{tabular}{lcccc}
    \hline
    Feature & Syn. & Cou. & LDC & TT \\
    \hline
    \multicolumn{1}{l}{Comparison} \\
    \quad with prior art & $\bullet$ & & & \\
    \quad with reference & & $\bullet$ & $\bullet$ & $\bullet$ \\
    SBGK \& ESBGK & & $\bullet$ & & $\bullet$ \\
    2D & & & $\bullet$ & \\
    2D axisymmetric & & & & $\bullet$ \\
    Diffuse boundaries & & $\bullet$ & $\bullet$ & $\bullet$ \\
    Shear flow & & $\bullet$ & $\bullet$ & \\
    Thermal flow & & $\bullet$ & & $\bullet$ \\
    Particle cloning/deletion \hspace{-5mm} & & & & $\bullet$ \\
    Adaptive equilibrium & & & & $\bullet$ \\
    \hline
  \end{tabular}
\end{table}

We present results validating the VRBGK implementation in PICLas.
Table \ref{tab:test_matrix} offers an overview of tested features.

\subsection{Synthetic Benchmarks}
\label{sec:synthetic_tests}

\begin{figure}
  \centering
  \begin{tikzpicture}[
      node distance=0.8cm and 2.5cm,
      term/.style={rectangle, draw, rounded corners, minimum width=1cm, minimum height=0.8cm, text centered, font=\fontsize{9}{9}\selectfont},
      process/.style={rectangle, draw, minimum width=3.8cm, minimum height=0.8cm, text centered, font=\fontsize{9}{9}\selectfont},
      decision/.style={diamond, draw,  text width=1.75cm, text centered, font=\fontsize{9}{9}\selectfont, aspect=1.2},
      arrow/.style={thick, -Stealth},
      every edge quotes/.append style={font=\fontsize{9}{9}\selectfont}
  ]

      \node (start) [term] {Start};
      \node (dec1) [decision, below=0.5cm of start] {$\text{Batch} < M_{pre}$};
      \node (proc1) [process, below=0.75cm of dec1] {Draw $\text{Pois}(\mathbb{E}[N_{eq}])$ particles};
      \node (proc2) [process, below=1.1cm of proc1] {$W_i = 1 \cdot \frac{f_{eq}(c_i)}{f(c_i)}$};
      \node (proc3) [process, below=1.1cm of proc2, text width=3.5cm] {Accumulate \\ $\text{var}[N_{vr}], \, \text{var}[W]$};
      \node (dec2) [decision, right=1.5cm of dec1] {$\text{Batch} < M$};
      \node (proc4) [process, below=of dec2] {Draw $\text{Pois}(\mathbb{E}[N_{eq}])$ particles};
      \node (proc5) [process, below=of proc4, text width=3.5cm] {
          $\begin{cases}
            W_i = \frac{N_{eq}}{\mathcal{N}(N_{eq},\, \sigma[N_{vr}])} \frac{f_{eq}(c_i)}{f(c_i)} \\
            W_i = \mathcal{N}\!\left(1,\, \frac{\sigma[W]}{\sqrt{N}}\right) \frac{f_{eq}(c_i)}{f(c_i)}
          \end{cases}$
      };
      \node (proc6) [process, below=of proc5, text width=3.5cm] {
        Accumulate \\  $\mathbb{E}[T], \, \text{var}[T], \, \mathbb{E}[u], \, \text{var}[u]$
      };
      \node (end) [term, above=0.5cm of dec2] {End};

      \draw [arrow] (start) -- (dec1);
      \draw [arrow] (dec1) -- node[right] {Yes} (proc1);
      \draw [arrow] (proc1) -- (proc2);
      \draw [arrow] (proc2) -- (proc3);
      \draw [arrow] (proc3.west) -- ++(-0.3cm, 0) |- node [above, xshift=0.5cm, yshift=0.1cm, font=\fontsize{9}{9}\selectfont] {Batch++} (dec1.west);
      \draw [arrow] (dec1.east) -- node[above] {No} (dec2.west);

      \draw [arrow] (dec2) -- node[right] {Yes} (proc4);
      \draw [arrow] (proc4) -- (proc5);
      \draw [arrow] (proc5) -- (proc6);
      \draw [arrow] (proc6.east) -- ++(0.2cm, 0) |- node[above, xshift=-0.5cm, yshift=0.1cm, font=\fontsize{9}{9}\selectfont] {Batch++} (dec2.east);
 
      \draw [arrow] (dec2) -- (end);

  \end{tikzpicture}
  \caption{Flowchart depicting the two-stage synthetic benchmark.
  The first loop determines the variance of $\bar{N}_{vr}$ and $W$, then the second loop calculates the statistical properties of flow variables.
  $M, M_{pre}$ are the number of batches.}
  \label{fig:synthetic_benchmark_flow_chart}
\end{figure}
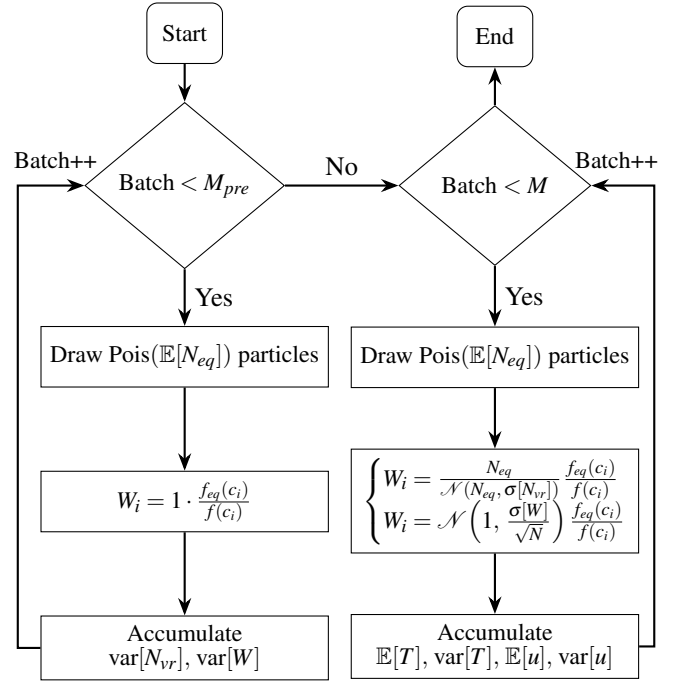

\begin{figure}
  \centering
  \includegraphics[width=\linewidth]{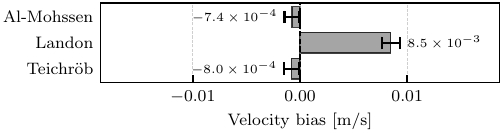}
  \caption{
      Bias in estimated velocity.
      Error bars indicate one standard deviation.
      Synthetic benchmark of methods for $u=\SI{15}{\meter\per\second}$, $u_{eq}=\SI{0}{\meter\per\second}$, $\E{N}=20$.}
  \label{fig:delta_u_bias_sim}
\end{figure}
\begin{figure}
    \centering
    \includegraphics[width=\linewidth]{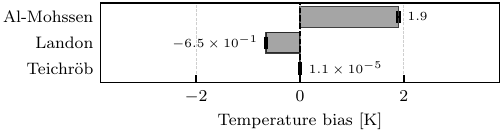}
    \caption{
      Bias in estimated temperature. 
      Error bars indicate one standard deviation.
      Synthetic benchmark of methods for $T=\SI{300}{\kelvin}$, $T_{eq}=\SI{400}{\kelvin}$, $\E{N}=200$.}
    \label{fig:delta_T_bias_sim}
\end{figure}
\begin{figure}
  \centering
  \includegraphics[width=\linewidth]{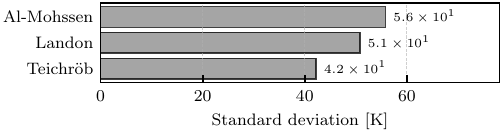}
  \caption{
      Standard deviation of estimated temperature.
      Synthetic benchmark of methods for $T=\SI{300}{\kelvin}$, $T_{eq}=\SI{400}{\kelvin}$, $\E{N}=200$.}
  \label{fig:delta_T_stddev_sim}
\end{figure}
\begin{figure}
  \centering
  \includegraphics[width=\linewidth]{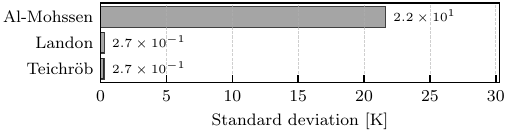}
  \caption{
      Standard deviation in estimated temperature for small deviation.
      Synthetic benchmark of methods for $T=\SI{300}{\kelvin}$, $T_{eq}=\SI{301}{\kelvin}$, $\E{N}=200$.}
  \label{fig:small_delta_T_stddev_sim}
\end{figure}

A synthetic benchmark is used to numerically confirm the estimator bias and variance analysis of Section \ref{sec:vr_formulation}.
As depicted in Figure \ref{fig:synthetic_benchmark_flow_chart}, the benchmark consists of two stages.
To produce realistically distributed weights, the first stage estimates the mean and standard deviation of $\bar{N}_{VR}$ and $\bar{W}$ in advance for use in the actual benchmarking stage.
For each batch, an ensemble of particles is generated according to a Maxwell-Boltzmann distribution $f(u, T)$ and with Poisson-distributed count.
In the second stage, particles are drawn identically, but the distribution of $\bar{N}_{VR}$ and $\bar{W}$ is used to produce more realistic weights $W_i$.
The estimate of velocity and temperature produced by each method for the current ensemble is recorded for analysis until the desired confidence interval has been achieved. 

We consider estimators according to Landon (Eq. \ref{eq:landon_estimator}), Al-Mohssen (Eq. \ref{eq:almohssen_estimator}), and this work (Eq. \ref{eq:R_vr}).
Figure \ref{fig:delta_u_bias_sim} and Figure \ref{fig:delta_T_bias_sim} reveal the bias present in Landon's VRBGK estimator in the presence of deviations from equilibrium (in $u$ and $T$ respectively).
The estimator according to Al-Mohssen presents with bias in the temperature but not in the velocity.
Both estimators are affected by a positive bias due to the overestimation of $N_{eq} / \bar{N}$, however only the $\bar{N}_{VR}$-based estimators also include additional bias from the calculation of the sample mean using $\bar{N}_{VR}$.
Because the equilibrium velocity is zero for the velocity bias measurement, overestimation of $N_{eq} / N$ does not express itself in the result.
The newly proposed estimator has a bias of \SI{0.000 \pm 0.019}{\kelvin} in temperature and \SI{-0.80 \pm 0.70}{\milli\meter\per\second} in velocity.
Therefore, there is no statistically significant bias at this simulation fidelity.

Under the previous conditions the standard deviation among all methods is comparable. 
On the other hand, observing the nominal case of small deviations from equilibrium in Figure \ref{fig:small_delta_T_stddev_sim} reveals a non-dimishing variance when using $N_{eq} / \bar{N}$ according to Al-Mohssens method.
Increased particle numbers can obscure the effect, as in Figure \ref{fig:small_delta_T_stddev_sim}, but eventually disturb the $\sigma = \mathcal{O} ( \E{R} - \E[f_{eq}]{R} )$ scaling.

\subsection{Couette}
We consider a steady, planar Couette flow.
The boundaries are infinite, fully diffuse walls at $x=0$ and $x=\SI{1}{\meter}$ with opposing velocities of $u_{\textrm{wall}} = \pm \SI{50}{\meter\per\second}$.
The grid is a linear, one-dimensional mesh of 100 cells. 
Initial conditions are stationary Argon at $\SI{280}{\kelvin}$ and $n=\SI{1.3722e19}{\per\meter\cubed}$.
A Knudsen number of $0.1$ is selected to demonstrate the utility of VR-BGK methods in the presence of wall-slip and non-equilibrium phenomena.
Sampling begins after the first $\SI{50}{\milli\second}$ and ends at $t_{end} = \SI{5}{\second}$ with \num{1e6} simulation particles for BGK, and \num{5e6} for BGK-DSMC and DSMC.

\begin{figure*}
    \begin{subfigure}{0.33\linewidth}
      \includegraphics[width=\linewidth]{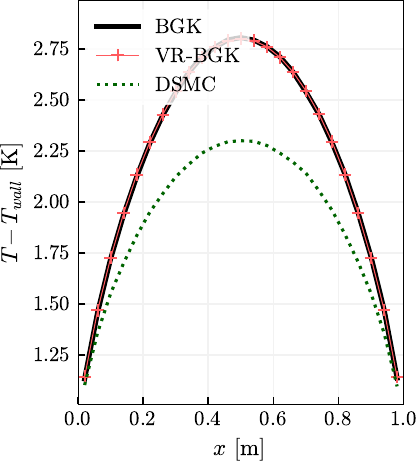}
      \caption{BGK}
      \label{fig:couette_bgk_T}
    \end{subfigure}
    \hfill
    \begin{subfigure}{0.33\linewidth}
      \includegraphics[width=\linewidth]{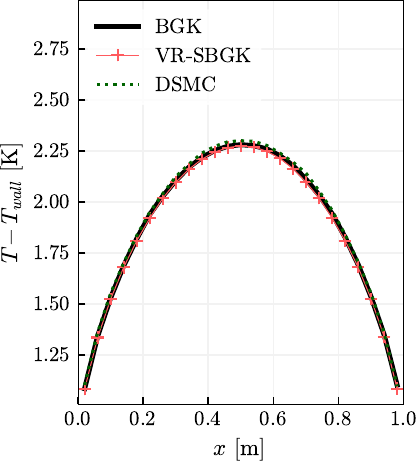}
      \caption{SBGK}
      \label{fig:couette_sbgk_T}
    \end{subfigure}
    \hfill
    \begin{subfigure}{0.33\linewidth}
      \includegraphics[width=\linewidth]{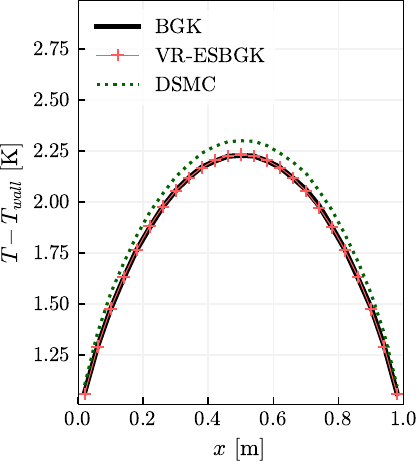}
      \caption{ESBGK}
      \label{fig:couette_esbgk_T}
    \end{subfigure}
    \caption{Temperature profiles for the Couette flow using different BGK models.}
    \label{fig:couette_T_profiles}
\end{figure*}

\begin{figure}
    \centering
    \includegraphics[width=0.9\linewidth]{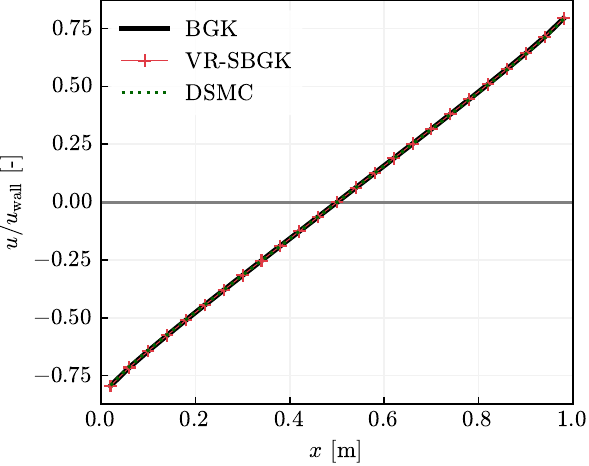}
    \caption{
      Comparison of the normalized Couette velocity profile for VR-SBGK and SBGK.}
    \label{fig:couette_u_profile}
\end{figure}

The selected wall velocity is low enough for the variance-reduced methods to not suffer from increased variance, yet high enough for a comparative simulation with BGK-DSMC and DSMC to be possible.
In addition, the viscous heating profile between the plates serves as a canonical test for correctness of the Prandtl number 
Alongside classical BGK, two extensions to the BGK method, SBGK and ESBGK, and their variance-reduced counterparts are simulated.

In Figure \ref{fig:couette_T_profiles}, the temperature profiles are plotted.
The agreement in temperature between all the variance-reduced and their non-variance-reduced variants is nearly exact.
The RMSE is $0.56\%$ for BGK, $1.20\%$ for SBGK and $0.55\%$ for ESBGK.
For this test case, comparing to the DSMC simulation as reference, SBGK yields the most accurate temperature profile.
All three methods are configurable in \mbox{PICLas} depending on the required application. 

A comparison of the velocity profile between SBGK and VR-SBGK is plotted in Figure \ref{fig:couette_u_profile}.
The agreement is excellent, with a RMSE of $0.020\%$.
Classic BGK and ESBGK give similar results of $0.030\%$ and $0.017\%$ RMSE respectively.

\subsection{Lid-Driven Cavity}
The lid-driven cavity problem investigates the validity and performance of the VRBGK method in two dimensions.
The cavity consists of a $\SI{1}{\meter} \times \SI{1}{\meter}$ square domain in the xy-plane with fully accommodating diffuse boundary conditions and an upper boundary at $y=\SI{1}{\meter}$ set to have a velocity of $\SI{10}{\meter\per\second}$ in the positive x-direction.
A $51 \times 51$ equidistant rectangular grid is used.
The solution is sampled during $\SI{0.025}{\second} < t < \SI{1}{\second}$ using $N = \num{5e5}$ particles for VRBGK and $N = \num{5e6}$ particles for BGK.
Due to the very low density and velocity of the Argon flow, the Reynolds number is also low ($Re \approx \SI{2e-10}{}$).
The resulting flow consists of a single, large central vortex~\cite{lid_driven_cavity}.
The Knudsen number is $Kn = 0.1$, leading to slip flow at the boundaries.

\begin{figure}
    \centering
    \includegraphics[width=0.9\linewidth]{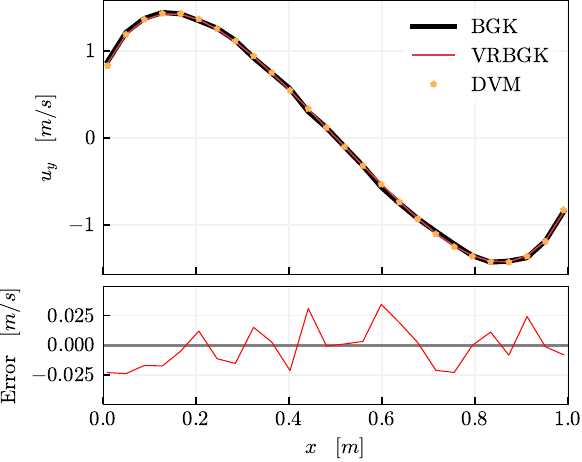}
    \caption{
        Velocity profile and error along x-axis at the midline $y=0.5$.
        The signed error is calculated as $u_{\text{VRBGK}} - u_{\text{BGK}}$.}
    \label{fig:cavity_y_velocity}
\end{figure}

\begin{figure}
    \centering
    \includegraphics[width=0.9\linewidth]{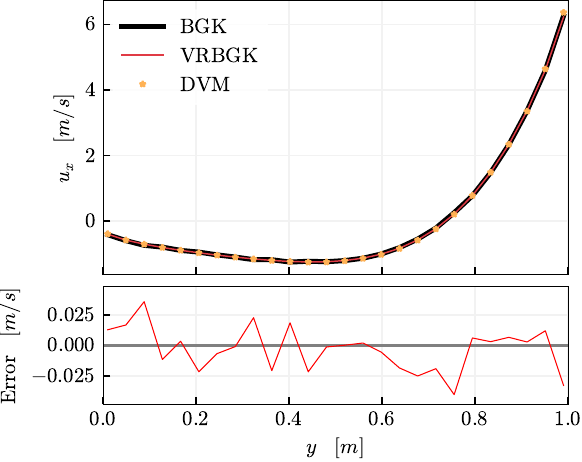}
    \caption{
        Velocity profile and error along y-axis at the midline $x=0.5$.
        The signed error is calculated as $u_{\text{VRBGK}} - u_{\text{BGK}}$.}
    \label{fig:cavity_x_velocity}
\end{figure}

Comparing the velocity profiles along the x- and y-centerlines, Fig. \ref{fig:cavity_y_velocity} and Fig. \ref{fig:cavity_x_velocity} show excellent agreement between the VRBGK and BGK methods.
The RMSE is $\SI{0.021}{\meter\per\second}$ for both the y-velocity and the x-velocity, which corresponds to a relative error of $0.21\%$.
Visually, the deviations do not show any bias and can largely be attributed to variance in the reference BGK simulation.
Due to the inefficiency of simulating in this regime for both VR \textit{and} non-VR methods, obtaining a much lower-noise reference case would be infeasible.

The Discrete Velocity Method (DVM) is another method of variance reduction, that has previously been validated and implemented in \mbox{PICLas}~\cite{felix_dvm}.
The agreement between DVM and the BGK methods is excellent as well.

\begin{figure}
    \centering
    \includegraphics[width=0.9\linewidth]{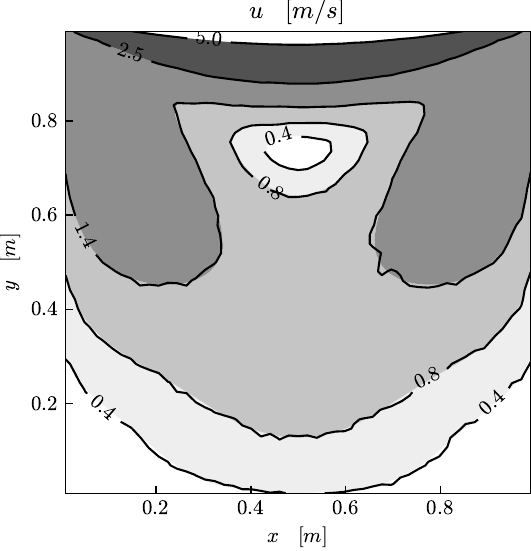}
    \caption{
      Comparison of velocity contours for VRBGK (filled contours) and BGK (contour lines).}
    \label{fig:cavity_contours}
\end{figure}

The runtime for the variance-reduced simulation is shortened by approximately a factor of 10 compared to BGK due to the reduced particle count.
In this case, the number of averaging steps remained unmodified while the particle count was tuned to achieve comparable variances in both methods, DVM excluded.
This would be completely synonymous with a change of the BGK solve time by increasing the number of averaging steps instead of the particle count, but the latter was preferred in the interest of maintaining a constant physical timescale.
A slower velocity flow would be required to showcase the algorithmic advantage of VRBGK.

\subsection{Thermal Transpiration}
\label{sec:thermal_creep}
\begin{figure}
    \centering
    \includegraphics[width=\linewidth]{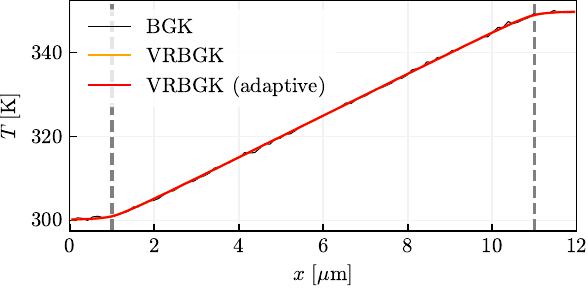}
    \caption{
        Temperature profile along the length of the micro-channel, averaged over $\SI{0.75}{\micro\second} < t < \SI{1}{\micro\second}$.}
    \label{fig:thermal_transpiration_T}
\end{figure}

Imposing a temperature gradient on a long, slender tube at high rarefaction induces a thermomolecular pressure to develop across the ends.
This effect is the working principle of Knudsen pumps and of interest for MEMS devices~\cite{KnudsenAReview}.
Because the magnitude of velocities and pressure differences in these applications can quickly become too small for classical particle methods, variance reduction can be beneficial. 
As a validation case, we select a circular \SI{1}{\micro\meter} diameter, \SI{12}{\micro\meter} long micro-channel.
A linear temperature gradient from \SI{300}{\kelvin} to \SI{350}{\kelvin} is applied across the length of the channel (Figure \ref{fig:thermal_transpiration_T}), excluding the first and last micrometer to avoid end-effects near the open boundaries.
To obtain the static pressure, an open end at \SI{1000}{\pascal}, modelled as a reservoir with $T=\SI{300}{\kelvin}$ and $n=\SI{2.414e23}{\per\meter\cubed}$, and a closed end modelled as a wall with $T=\SI{350}{\kelvin}$ are used.
All closed boundaries are fully accommodating.
The domain is modelled as a $\SI{12}{\micro\meter} \times \SI{0.5}{\micro\meter}$ grid with an axial symmetry condition along the long edge.
Both one-dimensional and two-dimensional grids were considered.
Simulations were carried out using ESBGK and \mbox{VR-ESBGK}, the latter with either global or adaptive equilibrium using an EMA smoothing factor of 0.95.
For the 2D grid, variable particle factors are used to enable efficient simulation in cells close to the symmetry axis.

An analytical solution for asymptotically long, slender channels according to \textcite{10.1116/1.579991} is available for comparison.
Sharipov's solution predicts a theoretical pressure difference of $\SI{73.16}{\pascal}$ in the absence of mass flow, however some error is expected due to the finite length of the simulated micro-channel.
The solution is computed using the tabulated values for $Q_P$ and $Q_T$~\cite{10.1116/1.579991}, which are within $0.5\%$ accuracy after interpolation.

\begin{figure}
    \centering
    \includegraphics[width=\linewidth]{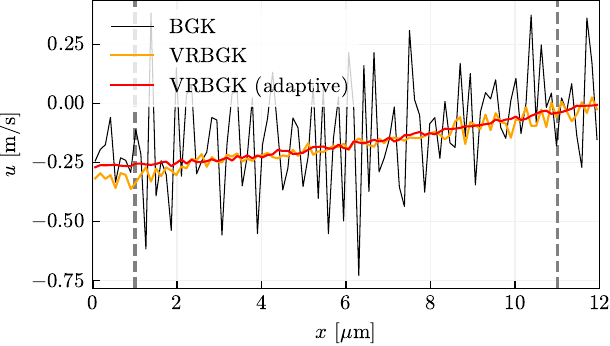}
    \caption{
        Velocity profile along the length of the micro-channel, averaged over $\SI{0.75}{\micro\second} < t < \SI{1}{\micro\second}$.
    }
    \label{fig:thermal_transpiration_v}
\end{figure}

Table \ref{tab:grid_study} shows that the one-dimensional grids consistently give results slightly greater than Sharipov's analytical method.
The pressure predicted by VRBGK on a $100 \times 1$ grid is \SI{74.46}{\pascal}, which is $1.8\%$ greater than the analytical result (Table \ref{tab:grid_study}) and thus in good agreement.
The 2D, $70 \times 5$ grid gives a result closer to the analytical result, however, this is not statistically significant due to the increased uncertainty in the result.
A plot comparing the flow velocity between BGK and VRBGK during the initial transient part of the solution from $\SI{0.75}{\micro\second}$ to $\SI{1}{\micro\second}$ is given in Figure \ref{fig:thermal_transpiration_v}.
For the non-adaptive VRBGK method, the median temperature and density of the ends are selected as a global equilibrium.
It is evident that in this regime, both VRBGK variants yield a significantly lower variance solution in the velocity than the BGK method.
In addition, it can be observed that the variance with adaptive equilibrium remains consistent throughout the domain despite the $\SI{50}{\kelvin}$ gradient, while the VRBGK method using a single global equilibrium state experiences increased variance at the ends of the micro-channel.
For this scenario, the global equilibrium approach is disadvantaged to have nearly equal variance in the pressure as the non-VR simulation.
This is due also to the moderately large variations in density.
Other quantities, such as the velocity (Figure \ref{fig:thermal_transpiration_v}), benefit more greatly from variance reduction.

\begin{table}
    \centering
      \caption{
        Results for the static pressure due to thermal transpiration with one standard deviation confidence.
        Measured at the closed end of the micro-channel averaged over $\SI{5}{\micro\second} < t < \SI{10}{\micro\second}$.\label{tab:grid_study}}
    \begin{tabular}{@{}p{3cm}p{1.5cm}l@{}}
        \toprule
        \textbf{Method} & \textbf{Grid} & \textbf{Static Pressure} \\
        \midrule
        VR-ESBGK & $50 \times 1$  & \SI{74.67 \pm 0.13}{\pascal} \\
        VR-ESBGK & $70 \times 1$  & \SI{74.82 \pm 0.07}{\pascal} \\
        VR-ESBGK & $100 \times 1$ & \SI{74.46 \pm 0.10}{\pascal} \\
        VR-ESBGK & $70 \times 5$  & \SI{73.99 \pm 0.35}{\pascal} \\
        \midrule
        VR-ESBGK (global) & $100 \times 1$ & \SI{72.37 \pm 0.82}{\pascal} \\
        ESBGK & $100 \times 1$ & \SI{72.64 \pm 0.99}{\pascal} \\
        \midrule
        Sharipov       & & \SI{73.16}{\pascal} \\
        \bottomrule
    \end{tabular}
\end{table}

The runtimes for the \SI{10}{\micro\second}, $100 \times 1$ thermal transpiration problem are \SI{595}{\second}, \SI{725}{\second}, and \SI{900}{\second} for ESBGK, \mbox{VR-ESBGK}, and \mbox{VR-ESBGK} with adaptive equilibrium respectively.
The cost per particle for \mbox{VR-ESBGK} with a global equilibrium is $22 \%$ greater than \mbox{ESBGK}.
When using adaptive equilibria, the additional local-to-global transformations and vice-versa increase the runtime cost by a further $24 \%$ over the global approach.
In light of the algorithmic advantage that VRBGK offers, no further optimization was deemed necessary.
The runtime of the baseline PICLas code prior to this work was \SI{663}{\second}.
The improvement in non-VR performance can be attributed to the use of an improved random number generator.

\section{Conclusion}

In this work, the Variance-Reduced BGK (VRBGK) method was implemented and advanced within the open-source gas-kinetics framework \mbox{PICLas}.
The theoretical foundation of the control variate approach was revisited, leading to the identification of limitations in prior density estimation schemes.
A new, unbiased estimator based on the sample mean of the VR-weights $\bar{W}$ was proposed and shown, through both analysis and synthetic benchmarks, to eliminate the systematic overestimation inherent in the approaches of Landon and Hadjiconstantinou, while also circumventing the non-diminishing variance affecting the VRDSMC method of Al-Mohssen for non-zero equilibrium moments.
The resulting method is both more stable and more amenable to mathematical analysis, particularly in the low particle number regime.

Results for a Couette flow, lid-driven cavity flow, and thermal transpiration problem demonstrate that the \mbox{PICLas} VRBGK implementation yields velocity, temperature, and density profiles in excellent agreement with non-VR reference cases.
The extension of the variance-reduced formulation to the Shakhov (SBGK) and Ellipsoidal Statistical (ESBGK) collision models was also demonstrated, with all three VR collision models giving results coincident with their non-VR counterparts.
This enables the correct simulation of flows where the correct Prandtl number is essential, such as those with combined shear and thermal effects.

The implementation into \mbox{PICLas} is ready for application to practical problems and can be enabled at runtime with only a small performance penalty.
The method is feature-rich, supporting diffuse and reflecting boundaries, variable particle factors, 2D axisymmetric simulations, and an adaptive equilibrium scheme that allows the reference state to vary across the domain, thus optimizing variance reduction for flows with large spatial gradients.

There are a number of opportunities for further development.
Support for mixed gas species would unlock further applications, such as simulating micro gas-chromatography systems~\cite{Zhao2025MonolithicIO} or atmospheric gases in MEMS devices.
By building on the conditional probability generalization of the control variate approach, most processes with a probabilistic description should be accessible to variance reduction.
Stability, however, remains a concern for the addition of new models and must potentially be counteracted with sufficient stabilization, for instance through kernel density estimation.

\subsection*{CRediT authorship contribution statement}
\textbf{Leon Teichröb}: Conceptualization, Methodology, Investigation, Software, Writing - Original Draft, Writing - Review \& Editing.
\textbf{Félix Garmirian}: Supervision, Writing - Original Draft, Writing - Review \& Editing.
\textbf{Marcel Pfeiffer}: Supervision, Writing - Review \& Editing, Software, Funding acquisition.

\subsection*{Data availability}
Our code is open source at \url{https://github.com/piclas-framework/piclas}

\subsection*{Declaration of competing interest}
The authors declare that they have no known competing financial
interests or personal relationships that could have appeared to influence
the work reported in this paper.

\subsection*{Acknowledgements}
This work is funded by the Deutsche Forschungsgemeinschaft (DFG, German Research Foundation) – Project-ID 516238647 – SFB 1667/1 (ATLAS - Advancing Technologies of Very Low-Altitude Satellites).

\printbibliography

\end{document}